\documentclass[12pt]{article}
\usepackage[cp1251]{inputenc}
\usepackage[russian, english]{babel}
\usepackage{cite}
\usepackage{amsmath}
\usepackage{amsfonts}
\usepackage{graphicx,color}
\def\d{\mathrm d}
\def\e{\mathrm e}

\def\th{\mathrm {th}}

\begin{document}

\begin{center}
{\Large\bf  Nonrelativistic quantum particles on the Minkowski plane}
\end{center}

\begin{center}
{\large N.\,A.~Gromov\textsuperscript{1},
I.\,V.~Kostyakov\textsuperscript{2}, 
V.\,V.~Kuratov\textsuperscript{3} }\\ [2mm]
{ Institute of Physics and Mathematics, \\ Komi Science Centre,  RAS,  Syktyvkar, Russia}
\\
\vspace{3mm}
{\textsuperscript{1}gromov@ipm.komisc.ru, %\\
\textsuperscript{2}kostyakov@ipm.komisc.ru, %\\
\textsuperscript{3}kuratov@ipm.komisc.ru
}
\end{center}
%\maketitle

\begin{abstract}

The quantum-mechanical problems of a nonrelativistic free particle, a harmonic oscillator and a Coulomb particle on Minkowski plane are discussed. The  Schr{\"o}dinger equations for eigenvalues  are obtained using the  Beltrami-Laplas operator of the pseudo-Euclidean plane and the corresponding potentials.  It is shown that, in contrast to the standard problem on Euclidean plane, in addition to the   continuous  spectrum, a free particle has a discrete energy levels and a Coulomb particle, in addition to the  discrete  spectrum, has unstable states that describe the incidence of a particle on isotropic lines forming a metric cone.

\end{abstract}

%%=================================================================================================================%%
%%----------------------------------------НАЧАЛО СТАТЬИ------------------------------------------------------------%%
%%=================================================================================================================%%

%\selectlanguage{russian}

 \section*{Introduction} 

With the development of nanophysics, it became possible to create new materials based on metaatoms, i.e., artificial structures of a more or less simple shape a few nanometers in size \cite{RK-18,D-19}. 
In particular, materials have been obtained that demonstrate the properties of a metal in one direction and behave like a dielectric in the orthogonal direction.
 They are called hyperbolic metamaterials \cite{Sm-12, Sm-15}.
Indeed, it is in spaces with a pseudo-Riemannian metric that there are two types of lines (besides isotropic) that are not compatible with isomorphisms and therefore allow you to model two different physical entities within the same space 
\cite{P-66, Gr-12, Gr-20}.
 
The rapid development of metamaterials in the past few years, the attractive prospects for their practical application in various fields stimulate theoretical studies of particle behavior in spaces with an unconventional metric.
In this paper, we consider three traditional precisely solvable problems on the energy levels of nonrelativistic quantum particles: a free particle, a harmonic oscillator, and Coulomb particles, in which the intrinsic space is a Minkowski plane. 
Previously, these problems were studied in \cite{GK-18,GKK-19,GKK-18}. 
Unlike the standard Euclidean plane problem, where the free particle has only a continuous spectrum, in the case of the Minkowski plane it additionally has discrete energy levels \cite{GK-18}.
In addition to the discrete spectrum, a Coulomb particle has unstable states describing the incidence of a particle on isotropic lines separating regions with positive and negative metrics \cite{GKK-19}.
 
It should be noted that the problem of a harmonic oscillator in relativistic space, whose Cartesian coordinates are interpreted as time and length, already appears in the theory of superstrings \cite{SGW,Kaku} %, \cite{Kaku} 
 and is usually called the "relativistic oscillator".
It has been known for quite some time and thoroughly worked out \cite{Bars}.
The Schr{\"o}dinger equation for eigenstates is usually solved
  in Cartesian coordinates and “non-physical” solutions are removed from the solutions (with a negative norm, unnormalized, etc.).
We consider the solution of this equation for an oscillator in the Minkowski plane in polar coordinates and we get three solutions that are usually not mentioned when solving in Cartesian coordinates 
(for $ M \neq 0 $).
All calculations are for one area only.
%All solutions are considered for only one area.
Note that solutions in polar coordinates for 1 + 1 harmonic oscillator for M = 0 are presented in \cite{Bars} 
(the complete solutions is given in Cartesian coordinates).
These tasks
were discussed extensively in the context of cosmology as well \cite{Bars1,Bars2,Neznamov3,Neznamov4}.

 We will start by studying the Coulomb particle. We obtain the results for a free particle at $ \alpha = 0 $.
For a quantum particle on the Euclidean plane, there is a dyon-oscillatory duality  \cite{Ter-Antonyan,GGT}.
A similar connection also exists on the Minkowski plane. Therefore, we obtain solutions for the harmonic oscillator from the corresponding solutions for the Coulomb particle.
 
%%%%%%%%%%%%%%%%%%%%%%%%%%%%%%%%%%%%%%%%%%%%%%%%% 

\section{Cartesian and polar coordinates on the Minkowski plane} 

\begin{figure}[h]
\begin{center}
\includegraphics[scale=0.8]{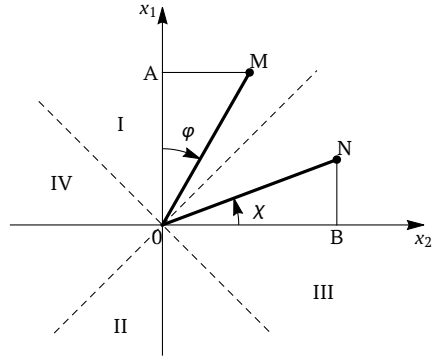}
\end{center}
  %\vspace{-6mm}
%\small \caption*{Рис.1. Полярные координаты $\{r, \varphi\}  $ точки M и координаты $\{\rho, \chi\} $ точки N на плоскости Минковского.
%OM=$r $, ON=$\rho $. \\ [1mm]
%Fig.1.  Polar coordinate $\{r, \varphi\},\; r=OM  $ of point M and coordinate $\{\rho, \chi\} $ of point N on Minkowski plane.} 
\caption{Polar coordinate $\{r, \varphi\},    $ of point M and coordinate $\{\rho, \chi\}   $ of point N on Minkowski plane. OM$=r,\;$ ON$=\rho $.}
 \label{ris1a} 
\end{figure}

The Minkowski plane is a two-dimensional space of zero curvature with a pseudo-Euclidean metrics $ s^2=x_1^2-x_2^2 $
 in Cartesian coordinates.
Cartesian and polar coordinates in the regions I $(x_1^2-x_2^2 >0,\; x_1>0)$ and II $(x_1^2-x_2^2 >0,\; x_1<0)$
 are related by the formulas (see fig. \ref{ris1a})

\begin{equation}
\left\{ \begin{array}{ll}
x_1= \pm r \ch \varphi     & \\
x_2= \pm r \sh \varphi  & \end{array}\right. 
\quad
\left\{ \begin{array}{ll}
{\tg r=\sqrt{x_1^2-x_2^2}>0},  & \\
{\th \varphi} =\displaystyle{{x_2}/{x_1}},   & \end{array}\right. 
\label{m1}
\end{equation}
where $r \in [0,\infty)$, $\varphi\in{\bf R}$.
%%%%%%%%%%%%%%%%%%%%%%%%%%%%%%%%%%%%%%%%%%%

In the regions III $(x_1^2-x_2^2 <0,\; x_2>0)$ and IV $(x_1^2-x_2^2 <0,\; x_2<0)$, instead of the imaginary $ r $, we introduce the real radius $ \rho = \sqrt{x_2^2-x_1^2}, $ i.e.
$ r = i \rho, $ and the new angle $ \chi, $ now counted from the axis $ x_2 $ and connected with the angle $ \varphi $ by the relation $\varphi=\chi -i\frac{\pi}{2} $, then
\begin{equation}
\left\{ \begin{array}{ll}
x_1=\pm \rho \,   \sh \chi &  \\
x_2=\pm \rho\, \ch \chi, & \end{array}\right. 
\quad
\left\{ \begin{array}{ll}
\th \rho ={\sqrt{x_2^2-x_1^2}}>0,    & \\
{\th \chi } =\displaystyle{{x_1}/{x_2}},      & \end{array}\right. 
\label{m2}
\end{equation}
where $\rho \in [0,\infty)$, $\chi\in{\bf R}$.

%%%%%%%%%%%%%%%%%%%%%%%%%%%%%%%%%%%%%%%%%%%%%%%%%%%%%%%%%%%%%%

\section{Schr{\"o}dinger equation }

The quantum-mechanical system on the Minkowski plane is described by the Schr{\"o}dinger equation
with the Hamiltonian $ H $, the kinetic part of which is proportional to the Beltrami-Laplace operator, and
 the potential $ U $ is described by a function of coordinates.
The Schr{\"o}dinger equation for the eigenvalues $H\Psi=E\Psi$ in Cartesian coordinates has the form
\begin{equation}
-\frac{\hbar^2}{2m}\left(\frac{\partial^2}{\partial x_1^2} -  \frac{\partial^2}{\partial x_2^2}  \right)\Psi(x_1,x_2) + U(x_1,x_2)\Psi(x_1,x_2) = E\Psi(x_1,x_2).
\label{ShEqM}
\end{equation} 
After moving to the polar coordinates (\ref{m1}), (\ref{m2})
the Hamiltonian is given by the expression
\begin{equation}
H=\left\{
\begin{aligned}
-\frac{\hbar^2}{2m}\left( \frac{\partial^2}{\partial r^2}+\frac{1}{r}\frac{\partial}{\partial r}
-\frac{1}{r^2}\frac{\partial^2}{\partial \varphi^2}\right)+
U(r,\varphi) \hspace{5mm} & \text{in the regions}\; I, \; II, \\
\frac{\hbar^2}{2m}\left( \frac{\partial^2}{\partial \rho^2}+\frac{1}{\rho}\frac{\partial}{\partial \rho}
-\frac{1}{\rho^2}\frac{\partial^2}{\partial \chi^2}\right)-
U(\rho,\chi) \hspace{5mm} & \text{in the regions}\; III, \; IV. \\
 \end{aligned}
\right.
\label{m3}
\end{equation}

We consider the Schr{\"o}dinger equation (\ref{ShEqM}) in polar coordinates in the regions $ I, II $
\begin{equation}
-\frac{\hbar^2}{2m}\left( 
\frac{\partial^2}{\partial r^2}+ \frac{1}{r}\frac{\partial}{\partial r}-
 \frac{1}{r^2}\frac{\partial^2}{\partial\varphi^2} \right)\Psi(r,\varphi) + U(r,\varphi)
 \Psi(r,\varphi) = E\Psi(r,\varphi).
\label{MPol}
\end{equation}
If we introduce the operator  $\hat{L}=-i\frac{\partial}{\partial \varphi}$,
 then this equation can be written as follows 
\begin{equation}
-\frac{\hbar^2}{2m}\left( 
\frac{\partial^2}{\partial r^2}+ \frac{1}{r}\frac{\partial}{\partial r}+
\frac{\hat{L}^2}{r^2} \right)\Psi(r,\varphi)%+
%$$
%\begin{equation}
+U(r)\Psi(r,\varphi) = E\Psi(r,\varphi).
\label{MinkPolShE2}
\end{equation} 
%%%%%%%%%%%%%%%%%%%%%%%%%%%%%%%%%%%%%%%%%%%%%
The eigenvalue $ M $ of the angular momentum operator $ \hat{L} $ corresponds to the solution
 \begin{equation}
   \hat{L}\Phi(\varphi)=M \Phi(\varphi), \quad \Phi(\varphi)= \frac{1}{\sqrt{2\pi}}\e^{i M\varphi},  
\label{L}
\end{equation} 
where $ M $, generally speaking, can be either real or complex number.
However, for purely imaginary values of $ M $, the states of the system will be unnormalizable.
For real values of $ M \in \mathbf{R} $, the angular part of the wave function is normalized to the delta function
\begin{equation}
  \int\limits_{-\infty}^{+\infty}\d \varphi \Phi(\varphi) \Phi(\varphi)^* =
\frac{1}{2\pi} \int\limits_{-\infty}^{+\infty}\d \varphi \, \e^{i(M-M')\varphi}=
  \delta(M-M').
\end{equation}
We are looking for a solution to the equation (\ref{MinkPolShE2}) in the form
\begin{equation}
\Psi(r,\varphi)=\frac{u(r)}{\sqrt{r}}\Phi(\varphi)
\label{L8}
\end{equation}
 with normalization of radial function
$u(r)\equiv  u_{n',M}(r) $ as follows
\begin{equation}
2\pi \int\limits_{-\infty}^{+\infty}\d r\, u^*_{n,M}(r)u_{n',M}(r)= \delta_{nn'},
\end{equation}
i.e. for wave function
\begin{equation}
\int |\Psi(r,\varphi)|^2 r\d r\d\varphi=   \delta(M-M')\delta_{nn'}.
\end{equation}
%%%%%%%%%%%%%%%%%%%%%%%%%%%%%%%%%%%%%%%%%%%%%%%%
%%%%%%%%%%%%%%%%%%%%%%%%%%%%%%%%%%%%%%%%%%%%%%%%%%
As a result, for the function $ u(r) $ we obtain the equation
\begin{equation}
 u''(r)+\left(\frac{2mE}{\hbar^2}+
 \frac{M^2+\frac{1}{4}}{r^2} - \frac{2m}{\hbar^2 }U(r)\right)u(r)=0.
 \label{uMF}
\end{equation}
Thus, the problem of the behavior of a quantum-mechanical system on the Minkowski plane was reduced to the one-dimensional Schr{\"o}dinger equation 
\begin{equation}
u''(r)+\frac{2m}{\hbar^2}\left(E -U_{eff}(r)\right) = 0
\label{L8q}
\end{equation}  
with the effective potential
\begin{equation}
% u''(r)+\frac{2m}{\hbar^2}\left(E -U_{eff}(r)\right) = 0, \quad
U_{eff}(r)= 
-\frac{\hbar^2}{2m}\frac{M^2+\frac{1}{4}}{r^2} + U(r).
\label{UeMink}
\end{equation}

\begin{figure}[!ht]
\begin{minipage}[b]{0.49\linewidth}
\begin{center}
\includegraphics[scale=0.8]{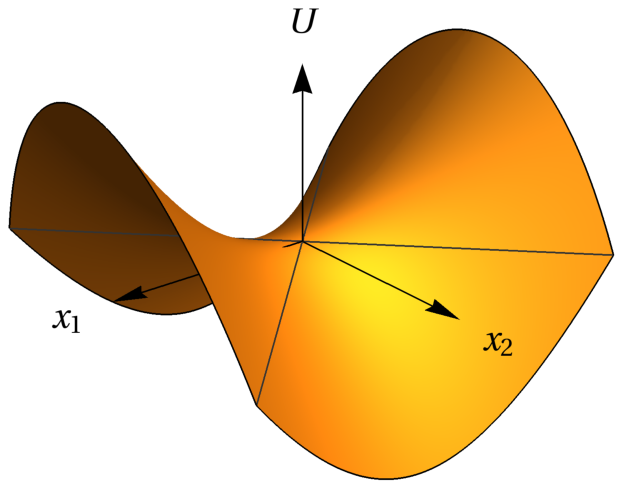} \\ %{ris4u.eps}
 \begin{center} %{ris4u.eps}
 a)
 \end{center}
\end{center}
\end{minipage}
\begin{minipage}[b]{0.49\linewidth}
\begin{center}
\includegraphics[scale=1.0]{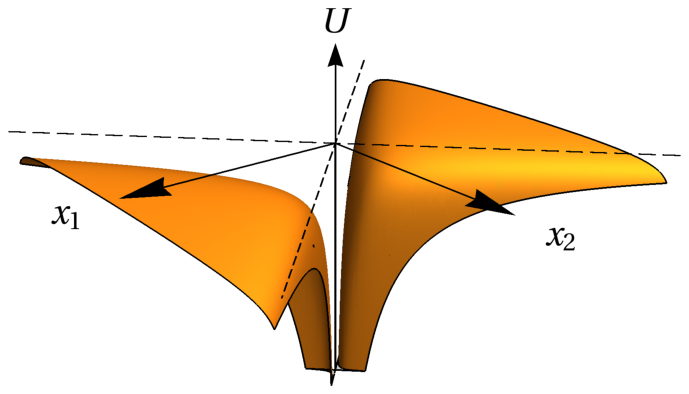}\\ %{ris6.eps}
\vspace{-15mm}
   \begin{center} %{ris4u.eps}
 b)
 \end{center}
\end{center}
\end{minipage}
\caption{Potentials (\ref{Pot}) of a harmonic oscillator a) and Coulomb particle b) on the Minkowski plane.}
 \label{ris4}
\end{figure}

\begin{figure}[h]
%\begin{figurehere}
\begin{minipage}[b]{0.32\linewidth}
\begin{center}
 \includegraphics[scale=0.7]{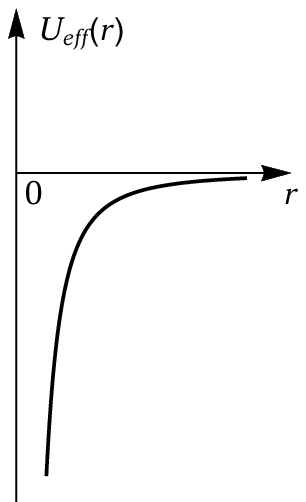}
 \begin{center} 
 a)
 \end{center}
 \end{center}
 \end{minipage}
\begin{minipage}[b]{0.32\linewidth}
\begin{center}
 \includegraphics[scale=0.7]{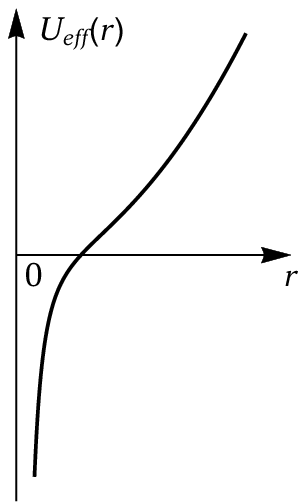}
 \end{center}
 \begin{center} 
 b)
 \end{center}
\end{minipage} 
\begin{minipage}[b]{0.32\linewidth}
\begin{center}
 \includegraphics[scale=0.7]{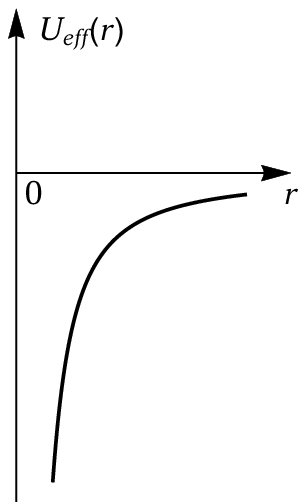}  %{QM_Mink_Coulomb_Ris-2.eps}
 \end{center}
 \begin{center} 
 c)
 \end{center}
\end{minipage} 
%\small \caption*{Рис.2. Эффективный потенциал кулоновской частицы на плоскости Минковского. \\ [1mm]
%Fig.2. Effective potential of Coulomb particle on Minkowski plane.}
\caption{Effective potentials (\ref{UeMink}), (\ref{Pot}) of a free particle a), harmonic oscillator b) and Coulomb particle c) on the Minkowski plane.}
 \label{ris8}
 %\end{figurehere}
 \end{figure}

%%%%%%%%%%%%%%%%%%%%%%%%%%
We consider in detail three precisely solvable problems: a free particle, a harmonic oscillator, and a Coulomb particle, for which the potential does not depend on the angle $ \varphi $
\begin{equation}
U(r)=\left\{ \begin{array}{ll}
 0, & \text{free particle} \\
\frac{1}{2}m\omega^2r^2,  & \text{oscillator} \\
-\frac{\alpha}{r}, & \text{Coulomb particle}  
\end{array}\right. .
\label{Pot}
\end{equation} 
The harmonic oscillator potential and the Coulomb potential are shown in Fig. \ref{ris4}, and the corresponding effective potentials (\ref{UeMink}) are shown in Fig. \ref{ris8}.

%%%%%%%%%%%%%%%%%%%%%%%%%%%%%%%%%%%%%%%%%%%5
%%%%%%%%%%%%%%%%%%%%%%%%%%%%%%%%%%%%%%%%%%%%%%%%

%%%%%%%%%%%%%%%%%%%%%%%%%%%%%%%%%%%%%%%%%%%%%%%%%%
%%%%%%%%%%%%%%%%%%%%%%%%%%%%%%%%%%%%%%%%%%%%%%%%%%%

\subsection{Coulomb particle}

We will start by studying the Coulomb particle. After replacing (\ref{L8}) from the Schr{\"o}dinger equation with the potential $ U(r) = - \frac{\alpha}{r} $
for the function $ u (r) $ from (\ref{uMF}) we have the equation
\begin{equation}
 u''(r)+\left(\frac{2mE}{\hbar^2}+
 \frac{M^2+\frac{1}{4}}{r^2}+\frac{2m\alpha}{\hbar^2 r}\right)u(r)=0.
 \label{uMFq}
\end{equation}
with boundary conditions
$ u(0)=u(\infty)=0.$ 
It appears in the relativistic equations for the Coulomb field and is analyzed, for example, in \cite{Perelomov,GTV}.
For small $ r $, the wave function $ u(r) $ behaves like  
\begin{equation}
%\sim \sqrt{r}\left(r^{iM}-\e^{-2i\gamma}r^{-iM}\right)
u(r) \sim\sqrt{r}\sin(M\ln r +\gamma),
\label{ur0}
\end{equation}
since in this case the equation takes the form
\begin{equation}
 u''(r)+ \frac{M^2+\frac{1}{4}}{r^2}\;u(r)=0.
\end{equation}
Its general solution is a superposition of particular  solutions 
  $r^{\frac{1}{2}+iM}$ and $r^{\frac{1}{2}-iM}$.
The parameter $ \gamma $ in (\ref{ur0}) is the phase of reflection from the singular cone $ r = 0 $.  
If $ \gamma $ is a complex number, then in the usual one-dimensional problem \cite{Perelomov} this means the presence of absorption at the point $ r = 0 $, and in our problem it can correspond to the passage of a particle through isotropic lines in regions III, IV in Fig. \ref{ris1a}.

A discrete spectrum is possible at negative energy values $ E $. It is convenient enter a unit of length
$ r_0 $, parameter $ g $ and dimensionless variable $ z $ by formulas
\begin{equation}
r_0=\frac{\hbar}{2\sqrt{-2mE}}, \quad g=\frac{ m \alpha}{\hbar\sqrt{-2mE}}, \quad z=\frac{r}{r_0}.
 \label{par}
\end{equation}
Then the equation (\ref{uMFq}) can be rewritten in the form
\begin{equation}
 u''(z)+\left( \frac{M^2+\frac{1}{4}}{z^2}+\frac{g}{z}-\frac{1}{4}\right)u(z)=0.
 \label{uMF2}
\end{equation}
For large $ z $, a solution decreasing at infinity behaves asymptotically as an exponential
  $u(z)\sim \e^{-\frac{z}{2}}$. Therefore, we can look for a solution to the equation (\ref{uMF2}) in the form
\begin{equation}
 u(z)=z^{\frac{1}{2}+iM}\e^{-\frac{z}{2}} f(z).
\end{equation}
For the function $ f(z) $ we obtain the equation
\begin{equation}
 zf''+\left(2iM+1-z\right)f'-\left(g+\frac{1}{2}+iM\right) f=0.
\end{equation}
Its basic solutions will be degenerate hypergeometric functions \cite{BE} of the first and second kind
$$ %\begin{equation}
f_1(z)=F\left(a,c,z \right)=1+\frac{a}{c}\frac{z}{1!}+ \frac{a(a+1)}{c(c+1)}\frac{z^2}{2!}+\ldots ,
$$
\begin{equation}
f_2(z)=\Phi\left(a,c,z \right)= z^{1-c}F\left(a-c+1,2-c,z \right),
\label{F}
\end{equation}
where  $c=1+2iM$, $a=iM+\frac{1}{2}-g$.

We have three types of solutions to the equation (\ref{uMF2})
$$%\begin{equation}
 u_1(z)=C_1\e^{- \frac{z}{2}}\sqrt{z}\;z^{iM} F\left(iM+\frac{1}{2}-g,2iM+1,z \right),
$$%\end{equation}
$$%\begin{equation}
u_2(z)=C_2\e^{-\frac{z}{2}}\sqrt{z}\;z^{-iM} F\left(\frac{1}{2}-iM-g,1-2iM,z \right),
$$%\end{equation}
\begin{equation}
u(z)=C_1u_1(z) +C_2u_2(z)= C\left(u_1(z) -\e^{-2i\gamma}u_2(z)\right) .
\label{u_1+u_2}
\end{equation}

Consider the first solution. For the function $ u_1(z) $ to tend to zero for large $ z $, it is necessary that the series 
(\ref{F}) break off. This is possible with $ a = -n, \; n = 0,1,2 \ldots $, then the eigenvalues
\begin{equation}
 E(n,M)=-\frac{m\alpha^2}{2\hbar^2\left(n+iM+\frac{1}{2}\right)^2},
 \label{E_NMink}
\end{equation}
and the wave function
\begin{equation}
 \Psi_1(z,\varphi)=C_1\e^{- \frac{z}{2}}\,z^{iM} F\left(-n,2iM+1,z \right) \e^{iM\varphi}.
 \label{Psi_1}
\end{equation}
For real $ M \neq 0 $, we have unstable decay states radiating to the singular center according to the interpretation of 
\cite{Shabad}.  
In our case, this corresponds to absorption by an isotropic cone or passage from regions I, II in regions III, IV in 
Fig. \ref{ris1a}.
 For $ M = 0 $, we obtain singlet states with a discrete spectrum, the same as on the Euclidean plane with $ M = 0 $, which is completely natural, since in these cases the particle moves only in the radial direction.
However, since the equation (\ref{MPol}) with the Coulomb potential is invariant under the replacement $ r \rightarrow -r $, the particle passes from region I to region II in Fig. \ref{ris1a}, in contrast to the Euclidean plane, where it is absorbed by the center.
 
It is easy to see that the first and second solutions (\ref{u_1+u_2}) are connected by simply replacing $ M $ with $ -M $,
therefore, the second solution is simply the complex conjugation of the first
$u_2(z)=u_1^*(z) $.

Consider the third solution $ u(z) $, presented as a superposition of the first two (\ref{u_1+u_2}).
Now it is already impossible to cut of  the series at $ u_1(z) $ and $ u_2(z) $ at the same time.
Let us pay attention to the behavior of this solution for large and small values of the argument.
When the $ z $ argument tends to zero, the wave function behaves like \cite{Perelomov}
 \begin{equation}
  u \sim  \sqrt{z}\left(z^{iM}-\e^{-2i\gamma}z^{-iM}\right)
     \sim \sqrt{r}\sin(M\ln r +\gamma-M\ln r_0).
 \end{equation}
 Phase $ \beta = \gamma - M\ln r_0 $
 depends on the energy $ \beta = \beta(E) $ and you can use the quantization condition \cite{Perelomov}
  \begin{equation}
 \beta(E_n)-\beta(E_0)=\pi n.
 \label{alpha_E_n}
\end{equation}

The asymptotic behavior of the functions $ u_1(z) $ and $ u_2(z) $ for $ z \rightarrow \infty $ has the form \cite{BE}
\begin{equation}
 u_1(z)\sim \e^{\frac{z}{2}}\;z^{-g} \frac{\Gamma\left(1+2iM\right)}{\Gamma\left(\frac{1}{2}+iM-g\right)}, \quad
 u_2(z)\sim \e^{\frac{z}{2}}\;z^{-g} \frac{\Gamma\left(1-2iM\right)}{\Gamma\left(\frac{1}{2}-iM-g\right)},
 \label{asimpt}
\end{equation}
then the solution $ u(z) $ at infinity behaves as follows
 \begin{equation} 
 u(r) \sim
  \e^{\frac{z}{2}}\;z^{-g} 
 \left(\frac{\Gamma\left(1+2iM\right)}{\Gamma\left(\frac{1}{2}+iM-g\right)} %- \right. 
%\left. 
-\e^{-2i\gamma}
 \frac{\Gamma\left(1-2iM\right)}{\Gamma\left(\frac{1}{2}-iM-g\right)}\right).
\end{equation}
For the function to go to zero, you need to require
 \begin{equation}
 \frac{\Gamma\left(1+2iM\right)}{\Gamma\left(\frac{1}{2}+iM-g\right)}=\e^{-2i\gamma}
 \frac{\Gamma\left(1-2iM\right)}{\Gamma\left(\frac{1}{2}-iM-g\right)}.
 \label{gamma}
\end{equation}
The quantization condition (\ref{alpha_E_n}) takes the form
\begin{equation}
 \gamma(E_n) - M \ln r_0(E_n)
= \gamma(E_0)-M \ln r_0 (E_0)+\pi n
\label{f}
 \end{equation}
 or 
\begin{equation}
 f(E_n) = f(E_0)+\pi n,
 \label{fEn}
 \end{equation}
 where 
\begin{equation}
 f(E)=-M \ln g(E) + \mathrm{arg} \frac{\Gamma\left(\frac{1}{2}-g(E)+iM\right)}{\Gamma\left(1+2iM\right)}.
 \label{fE}
\end{equation}
 
 For large negative energy values (i.e., for small $ r $) and $ M \neq 0 $, the function $ f(E) \approx -M \ln g(E) $. Then for energy levels we get the same formula
 \begin{equation}
  E_n=E_0 \e^{ \frac{2\pi n}{M}}, \quad  E_0<0, \quad n=0, 1,  2,\ldots, 
  \label{EN_f}
 \end{equation}
as in \cite{Perelomov}.
Discrete energy levels of a Coulomb particle become more rare with $ E \rightarrow - \infty $.
Indeed, in this case, the effective potential is approximately equal to
\begin{equation}
U_{\text{eff}}(r)\approx
-\frac{\hbar^2}{2m}\frac{M^2+\frac{1}{4}}{r^2},
\end{equation}
i.e. is determined only by orbital forces.
In \cite{Neznamov}, such
the solutions are called the "event horizon fall mode".
Solution of the one-dimensional Schr{\"o}dinger equation
with potential $ U = - \beta / r^2 $ and the appearance of discrete energy levels
was discussed in \cite{LL,Perelomov,GTV} and was
confirmed in \cite{Neznamov2} by numerical experiment.

For negative energies tending to zero (large values of $r $ and $ g(E) $), the function
$ f(E) \approx - \pi g(E) $
 and energy levels are given by the formula
\begin{equation}
 E_n=-\frac{m\alpha^2}{2\hbar^2\left(n+g_0\right)^2},\quad n=0,1,2,\dots 
 \label{E_n}
\end{equation}
For $g_0=\frac{1}{2}$  we obtain a formula that describes the energy levels of the Coulomb particle on the Euclidean plane.
Thus, when moving away from the isotropic cone, levels are condensed, just like in the ordinary Coulomb potential when moving away from the attracting center. Indeed, in both cases, the Coulomb potential is the leading one in this energy region.
Since for large values of $ g(E) $ the quantum numbers $ n $ are also large ($ n \gg g_0 $),
then for energy levels in this area we  can approximately write
\begin{equation}
 E_n=-\frac{m\alpha^2}{2\hbar^2 n^2}.
 \label{E^n}
\end{equation}

\subsection{Free particle}

The energy levels for a free particle on the Minkowski plane can also be obtained directly from
formulas (\ref{fEn}), (\ref{fE}) for $ \alpha = 0 $ (absence of the Coulomb potential).
In this case
\begin{equation}
 f(E)=-M \ln r_0(E) + C(M),
\end{equation}
where $ C(M) $  is a function of $ M $.
Substituting this expression in (\ref{fEn}), we obtain the formula (\ref{EN_f}),
coinciding with previously obtained energy levels for a free particle \cite{GK-18}.
With increasing $ n $, the distance between the levels increases and the discrete spectrum at
$ E \rightarrow - \infty $ is becoming increasingly rare.
For positive energies $ E> 0 $, the spectrum is continuous.

\subsection{Harmonic oscillator}

The description of the harmonic oscillator on the Minkowski plane is obtained from the description of the Coulomb particle
using the analogue of dion-oscillatory duality, which takes place on the Euclidean plane \cite{Ter-Antonyan,GGT}.
If in the equation (\ref{MPol}) with the potential $ U(r) = - \frac{\alpha}{r} $ and the eigenvalue $ E_C $ make a change of variables
$ r_0r = \varrho^2 $, $ \varphi = 2 \phi $, where $ r_0 $  is the scale factor,
having a dimension of length, we obtain the Schr{\"o}dinger equation for the oscillator
\begin{equation}
-\frac{\hbar^2}{2m}\left( 
\frac{\partial^2}{\partial \varrho^2}+ \frac{1}{\varrho}\frac{\partial}{\partial \varrho}-
\frac{1}{\varrho^2}\frac{\partial^2}{\partial\phi^2} \right)\Psi(\varrho,\phi)
+\frac{1}{2}m\omega^2\varrho^2\Psi(\varrho,\phi) = E_{osc}\Psi(\varrho,\phi),
\label{MinkPolShEOsc}
\end{equation} 
where $r_0E_{osc}=4\alpha$, $m\omega^2r_0^2=-8E_C$. 
Note that such a relativistic oscillator plays an important role in the theory of superstrings \cite{SGW,Kaku}.
%%%%%%%%%%%%%%%%
Thus, in order to go to the formulas for the oscillator in the corresponding formulas for solving the Coulomb problem, we need to make a substitution
\begin{equation}
 r_0r=\varrho^2, \quad \varphi=2\phi, \quad
 r_0E_{osc}=4\alpha, \quad  m\omega^2r_0^2=-8E_{C}, \quad 
 M_{osc}=2 M_C.
 \label{zamena}
\end{equation}
%%%%%%%%%%%%%%%%%%%%%%%%

As a result, the energy levels (\ref{E_NMink}) of the Coulomb particle for the first solution (\ref{u_1+u_2}) will go over
 into oscillator energy levels 
\begin{equation}
  E_{osc}=\hbar\omega\left(2n+1+iM_{osc}\right),
 \label{E_{osc}}
\end{equation}
and the wave function (\ref{Psi_1}) is converted into the wave function of the oscillator
\begin{equation} 
 \Psi_{osc}(\varrho,\phi)=C_{osc}\varrho^{iM_{osc}}
 \e^{-\frac{m\omega}{2\hbar}\varrho^2}
 F\left(-n, 1+iM_{osc};\frac{m\omega}{\hbar}\varrho^2\right)\e^{iM_{osc}\phi}.
 \label{psi_1}
  \end{equation}
This type of solutions for $ M \neq 0 $ on the Euclidean plane
are interpreted as solutions for particles,
falling on the center or outgoing from it.
In the case of the Minkowski plane, such solutions can
interpret not only as “disappearing” on the light
cone or “emerging” from it, but also as passing into the region of $ III $ or $ IV $. 
For $ M = 0 $, these solutions describe singlet,
positive-norm, Lorentz-invariant physical states
with discrete spectrum. For the second solution (\ref{u_1+u_2}) everything is similar.

The condition (\ref{gamma}) of the tendency of the third solution (\ref{u_1+u_2}) to zero with an unlimited increase in the argument can be rewrite as 
\begin{equation}
 \frac{\Gamma(1+iM_{osc})}{\Gamma\left(\frac{1+iM_{osc}}{2}-\frac{E_{osc}}{2\hbar\omega}\right)} =
 e^{-2i\gamma}\frac{\Gamma(1-iM_{osc})}{\Gamma\left(\frac{1-iM_{osc}}{2}-\frac{E_{osc}}{2\hbar\omega}\right)},
% \end{equation} 
%$$
\quad  g=\frac{ m \alpha}{\hbar\sqrt{-2mE_C}}=
 \frac{E_{osc}}{2\hbar\omega}.
%$$
 \end{equation}
 The quantization condition (\ref{fEn}) with the function (\ref{fE}) retains its form.
Given the relationship between the energies of the Coulomb particle and the oscillator 
\begin{equation}
E_{osc} =\pm \frac{2\alpha\omega \sqrt{m}}{\sqrt{-2E_C}},
\label{con}
 \end{equation}
which is easy to obtain from (\ref{zamena}), we have
at low positive energies (at low $ \varrho $) and $ M_{osc} \ne 0 $ oscillator spectrum
\begin{equation}
  E_n^{osc}=E_0^{osc} \e^{ \frac{2\pi n}{M_{osc}}}, \quad E_0^{osc}>0,\quad n=0,- 1, - 2,\ldots,  
  \label{EN_{fosc}}
 \end{equation}
 i.e. a formula similar to the case of a Coulomb particle (\ref{EN_f}).
Discrete oscillator energy levels are condensed when approaching the zero value. 
In the limit of large positive energies (large values of $\varrho $) from (\ref{E_n}) we obtain  
\begin{equation}
 E_{osc}=2\hbar\omega(n+g_0), \quad n=0, 1, 2, \ldots .
 \label{con1}
\end{equation}
For large quantum numbers ($ n \gg g_0 $), it takes a standard expression describing the discrete energy levels of a classical oscillator.

\section{Conclusion}

\begin{figure}[h]
%\begin{figurehere}
\begin{center}
 \includegraphics[scale=0.8]{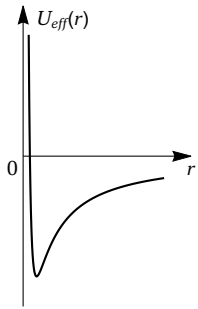}  %{QM_Mink_Coulomb_Ris-1.eps}
 \end{center}
%\small \caption*{Рис.3. Эффективный потенциал кулоновской частицы на евклидовой плоскости. \\ [1mm]
%Fig.3. Effective potential of Coulomb particle on Euclidean  plane.}
\caption{Effective potential  of a  Coulomb particle  on the Minkowski plane.}
 \label{ris3}
% \end{figurehere}
\end{figure}

Effective Coulomb potential \cite{LL} particles on the Euclidean plane (Fig. \ref{ris3})
\begin{equation}
U_{\text{eff}}(r)=
-\frac{\hbar^2}{2m}\frac{\frac{1}{4}-M^2}{r^2} - \frac{\alpha}{ r},
\label{UeEu}
\end{equation}
differs from the Coulomb potential (\ref{UeMink})
on the Minkowski plane only by a  sign at $ M ^ 2 $
(Fig. \ref{ris8}).

A similar sign-changing effect occurs in the relativistic Dirac and Klein-Gordon equations for
Coulomb potential \cite{Neznamov}, where  for particles with a charge $ Z $ less than a certain critical charge 
$ Z <Z_ {cr} $,  the Euclidean case is realized, and for particles with a charge greater than the critical $ Z> Z_{cr} $ --- the case of the Minkowski plane.

%%%%%%%%%%%%%%%%%%%%%%%%%%%%
 %

%%%%%%%%%%%%%%%%%%%%%%%%%%%%%

For a harmonic oscillator on the Minkowski plane, unstable decay states (\ref{psi_1}),
arising for real $ M \neq 0 $ are interpreted not only as "disappearing" $ \, $ on an isotropic
cone or “appearing” $ \, $ from it, but also as passing from regions I, II to regions III, IV in Fig. \ref{ris1a}.
For $ M = 0 $, we obtain the same states with a discrete spectrum as on the Euclidean plane, which is completely natural,
since in these cases the particle moves only in the radial direction.
However, since the equation (\ref{MPol}) with the Coulomb potential is invariant under the replacement $ r \rightarrow -r $, the particle passes from region I to region II in Fig. \ref{ris1a}, in contrast to the Euclidean plane, where it is absorbed by the center.

Thus, although solutions of the equations (\ref{uMF}), (\ref{L8q}) with potentials (\ref{Pot}) have long been known, but they arose in other problems. In problems of nonrelativistic quantum particles on the Minkowski plane, these solutions receive a new physical interpretation.

\end{document}